\documentclass[pra,twocolumn,floatfix,a4paper,superscriptaddress]{revtex4}
\usepackage{amssymb}
\usepackage{amsfonts}
\usepackage{bm,color,amsmath,txfonts}
\usepackage{graphicx}
\usepackage{epstopdf}
\usepackage{siunitx}
\usepackage{subfigure}
\usepackage{verbatim}
\usepackage{dcolumn}
\usepackage{bm}
\usepackage{epsf}
\usepackage{xcolor}
\usepackage{hyperref}
\usepackage{hhline}

\usepackage{float}
\usepackage{enumerate}
\usepackage{bbm}
\usepackage{lipsum}
\usepackage{mathrsfs}
\begin{document}
	\title[Short Title]{Generating entanglement of two acoustic modes by driving the qubit in circuit quantum acoustodynamics system}
	\author{Mei-Rong Wei}
	\affiliation{State Key Laboratory of Quantum Optics and Quantum Optics Devices, Institute of Opto-Electronics, College of Physics and Electronic Engineering, Shanxi University, Taiyuan, Shanxi 030006, China}
	\affiliation{Collaborative Innovation Center of Extreme Optics, Shanxi University, Taiyuan 030006, China}
	\author{Qi Guo\footnote{E-mail: qguo@sxu.edu.cn}}
	\affiliation{State Key Laboratory of Quantum Optics and Quantum Optics Devices, Institute of Opto-Electronics, College of Physics and Electronic Engineering, Shanxi University, Taiyuan, Shanxi 030006, China}
	\affiliation{Collaborative Innovation Center of Extreme Optics, Shanxi University, Taiyuan 030006, China}
	\author{Gang Li}
	\affiliation{State Key Laboratory of Quantum Optics and Quantum Optics Devices, Institute of Opto-Electronics, College of Physics and Electronic Engineering, Shanxi University, Taiyuan, Shanxi 030006, China}
	\affiliation{Collaborative Innovation Center of Extreme Optics, Shanxi University, Taiyuan 030006, China}
	\author{Tiancai Zhang\footnote{E-mail: tczhang@sxu.edu.cn}}
	\affiliation{State Key Laboratory of Quantum Optics and Quantum Optics Devices, Institute of Opto-Electronics, College of Physics and Electronic Engineering, Shanxi University, Taiyuan, Shanxi 030006, China}
	\affiliation{Collaborative Innovation Center of Extreme Optics, Shanxi University, Taiyuan 030006, China}
\begin{abstract}
		
We propose how to generate the entanglement of two long-lived phonon modes in a circuit quantum acoustodynamics system, which consists of a multi-mode high-frequency bulk acoustic wave resonator and a transmon-type superconducting qubit. Two acoustic modes couple to the qubit through piezoelectric interaction, and the qubit is driven by a microwave field. Under the condition of far detuning between the qubit and acoustic modes, the qubit can be eliminated adiabatically, and thus establishing the indirect interaction between the two acoustic modes. We demonstrate that such the indirect interaction can be the parametric-amplification-type interaction by appropriately choosing the drive frequency and strength, so the entanglement between acoustic modes can be created by the direct unitary evolution. We numerically analyze the parameter conditions for generating the entanglement in detail and evaluate the influence of system dissipations and noise. The results show that the scheme can be realized using currently available parameters and has strong robustness against the dissipations and environmental temperature. This work may provide efficient resource for the quantum information processing based on the phononic systems.
		
\end{abstract}
\maketitle
	
\section{Introduction}
	
Quantum entanglement is one of the most striking phenomena in quantum mechanics, describing non-classical correlations within quantum systems \cite{1}. Since the renowned Einstein-Podolsky-Rosen paradox was proposed by Einstein, Podolsky, and Rosen, quantum entanglement has remained a focal point in quantum physics, as it is not only a fundamental phenomenon within quantum mechanics but also crucial for the development of quantum technologies \cite{2,3}. With the ongoing advancement of theoretical and experimental techniques for quantum state preparation, physicists have successfully utilized a variety of physical systems, such as cavity optomechanical systems \cite{4,5,6,7,8,9,10,11,12,13,14,14a}, magnonic hybrid systems \cite{15,16,17,18,19,20,21,22}, superconducting circuits systems \cite{22a,23,24,25,25a}, and circuit quantum acoustodynamics (cQAD) systems \cite{26,27,28,29,30,30a,30b}, to achieve the preparation and manipulation of macroscopic quantum states, despite the numerous complex factors constrain the observation of quantum effects in macroscopic objects.
	
In recent years, high-frequency bulk acoustic wave resonator (HBAR), which can support specific acoustic modes of gigahertz-frequency standing waves, exhibits unique advantages among numerous physical systems due to its potential to integrate with superconducting qubits through linear piezoelectric interactions \cite{31,32,33,34,35,36,37,38,38a,38b}. Meanwhile, HBAR is capable of producing low-loss phonon sources with loss rates in the $\rm kHz$ regime and the coherence time in the order of hundreds of microseconds \cite{26,27}. The cQAD system, a novel hybrid system comprising of the HBAR and a superconducting qubit, enables the complex quantum control of macroscopic mechanical objects, thereby opening up the potential for utilizing acoustic modes as quantum resources \cite{26}. Many notable experimental advances based on cQAD system have been achieved, including direct measurements of phonon number distribution and parity of non-classical mechanical states in the strong dispersive regime \cite{27}, as well as the preparation of the Sch{\"o}rdinger cat state in a 16-microgram mechanical oscillator in the resonant coupling regime \cite{28}. Very recently, von L{\"o}pke \emph{et al}. demonstrated the Hong-Ou-Mandel effect between phonons by engineering the beam-splitter-type interaction between two mechanical modes \cite{29}, and Marti \emph{et al}. proposed a scheme for phonon squeezing by using two-tone driving of a qubit \cite{30}. These developments indicate that the cQAD system can serve as an ideal platform for quantum acoustics, and offers considerable experimental support for the study of macroscopic quantum states. Therefore, we here focus on the generation of entanglement for solid-state phonon modes based on cQAD system, which, to our knowledge, has been almost unexplored to date.
	
There have been numerous methods for preparing quantum states in hybrid systems, such as introducing auxiliary systems \cite{10}, incorporating nonlinear crystals \cite{11}, quantum feedback control\cite{39}, amplitude modulation of the driving field \cite{9,10}, and the reservoir-engineering method \cite{5,6,7}. However, these methods typically increase the complexity of the system and require more advanced technical requirements. In comparison, the approach based on the unitary dynamics of the system is simpler and more direct for generating quantum states, especially for cQAD systems with long coherence duration. Some methods based on unitary evolution have been developed in cavity quantum electrodynamics system \cite{40} and circuit electrodynamics system \cite{40b}, and these methods have also been extended to the cavity magnonics system \cite{18,20}. However, the direct interactions between two acoustic modes in HBAR is still a challenge. Based on the aforementioned works, in this paper, we construct the indirect interaction between the two acoustic modes in cQAD system by eliminating the qubit adiabatically under the condition of far-detuning phonon-qubit coupling. What's exciting is that the parametric-amplification-type interaction between the two acoustic modes can be obtained by driving the qubit with appropriate frequency and strength, which means the two-mode squeezing (i.e., entanglement) of the acoustic modes can be generated by the unitary evolution of the system. We also show the acoustic-mode entanglement under different system parameters, including the detuning between the qubit and drive field, the driving strength, and the phonon-qubit coupling strength. It is demonstrated that the scheme is feasible under the current experimental conditions and robust against system dissipations and environmental temperature.
	
The organization of this paper is as follows. In Sec.$~$II, we describe the basic model of the scheme and derive the effective Hamiltonian of the system. In Sec.$~$III, we present numerical results of the entanglement evolution, show the consistency of entanglement obtained from the effective Hamiltonian and the full Hamiltonian; analyze the effects of various system parameters on the long-lived entanglement, and demonstrate the robustness of the entanglement against thermal noise. Finally, conclusions are presented in Sec.$~$VI.
	
\section{Theoretical model and the effective Hamiltonian}
	
\begin{figure}
	\centering
	\includegraphics[width=\linewidth]{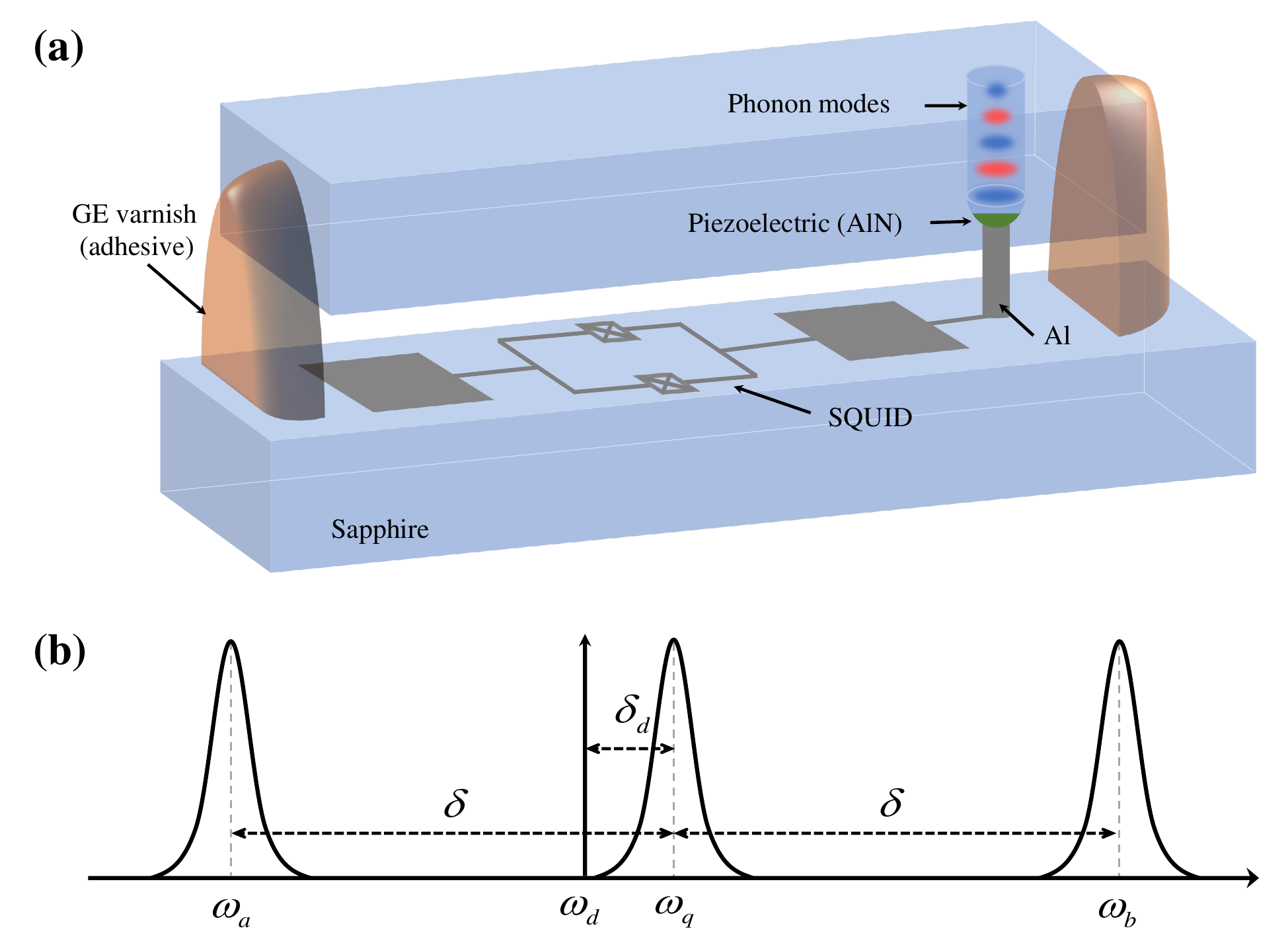}
	\caption{(a) Schematic of the circuit quantum acoustodynamics system. The HBAR chip, positioned at the top, is equipped with a layer of piezoelectric aluminum nitride (green) and supports two acoustic modes (blue and red). The superconducting transmon qubit on the lower chip couple to the HBAR through piezoelectric interaction. (b) Frequency spectrum of the system. The qubit with frequency $\omega_q$ is driven by a microwave field with frequency $\omega_d$, where $\delta_d=\omega_q-\omega_d$.}
\end{figure}
	
The cQAD system utilized in the scheme is shown in Fig.~\ref{1}(a), where the superconducting transmon qubit and the HBAR are individually fabricated on a sapphire substrate and then bonded together with small drops of GE varnish applied to the edges \cite{26,28,29,31}. The semicircle beneath the HBAR is designed to suppress the generation of transverse waves. The HBAR has multi-mode properties where two acoustic modes are coupled to the electric field of the qubit via the piezoelectric transducer (AlN), as detailed in Ref.~\cite{29}. Concurrently, the device is housed in a three-dimensional aluminum cavity to protect the qubit from its environment \cite{28}. The Hamiltonian (in unit of $\hbar= 1$) of our system is:
\begin{align}\label{1}	
	H=&\omega_{a} a^{\dagger} a+\omega_{b} b^{\dagger} b+\frac{\omega_{q}}{2} \sigma_{z}+g_{a}(\sigma_{e g} a+\sigma_{g e} a^{\dagger})\notag\\
	&+g_{b}(\sigma_{e g} b+\sigma_{g e} b^{\dagger})+\Omega_d(e^{-i \omega_{d} t} \sigma_{e g}+e^{i \omega_{d} t} \sigma_{g e}).
\end{align}
Here, $a$ $(a^\dagger)$ and $b$ $(b^\dagger)$ are the annihilation (creation) operators of the phonon modes with frequencies $\omega_{a}$ and $\omega_{b}$, respectively. Due to its strong anharmonicity, the superconducting circuit can be regarded as a two-level system (qubit) with transition frequency $\omega_q$, where $|g\rangle$ $(|e\rangle)$ is the ground (excited) state. The Pauli matrix is defined as $\sigma_z=|e\rangle\langle e|-|g\rangle\langle g|$, and $\sigma_{eg}=|e\rangle\langle g|$ $(\sigma_{ge}=|g\rangle\langle e|)$ is the raising (lowering) operator of the qubit. $g_j$ ($j=a,b$) denotes the coupling strength between the phonon and qubit. The qubit is driven by the microwave field with amplitude $\Omega_d$ and frequency $\omega_d$.
	
The Hamiltonian, in the interaction picture with the transformation $U_{1} =e^{-i( \omega_{a} a^\dagger a + \omega_{b} b^\dagger b + \frac{\omega_{q}}{2} \sigma_{z})t}$, can be expressed as follows
\begin{align}\label{2}	
	H_{1} =&g_a(\sigma_{e g} a e^{i \delta t}+\sigma_{g e} a^{\dagger} e^{-i \delta t})+g_b(\sigma_{e g} b e^{-i \delta t}+\sigma_{g e} b^{\dagger} e^{i \delta t})\notag\\
	&+\Omega_{d}(e^{i \delta_{d} t} \sigma_{e g}+e^{-i \delta_{d} t} \sigma_{g e}),
\end{align}
where $\delta=(\omega_{b}-\omega_a)/2$ (i.e., $\omega_q=(\omega_a+\omega_b)/2$) and $\delta_d=\omega_q-\omega_d$, as illustrated in Fig.~\ref{1}(b). In order to simplify the calculation, we write the Hamiltonian in the rotating frame of $U_{2}=e^{\frac{i}{2} \delta_{d} \sigma_{z} t}$ to eliminate the time factors in the driving term
\begin{align}\label{3}	
	H_{2}=&H_{0}^{\prime}+g_{a}[\sigma_{e g} a e^{i(\delta-\delta_{d}) t}+\sigma_{g e} a^{\dagger} e^{-i(\delta-\delta_{d})t}]\notag\\
	&+g_{b}[\sigma_{e g} b e^{-i(\delta+\delta_{d})t}+\sigma_{g e} b^{\dagger} e^{i(\delta+\delta_{d})t}],
\end{align}
where we define $H_{0}^{\prime}=\frac{\delta_{d}}{2} \sigma_{z}+\Omega_{d}(\sigma_{e g}+\sigma_{g e})$. To elucidate the underlying physical mechanisms, we diagonalize the Hamiltonian $H_{0}^{\prime}$ to obtain the dressed states \(|\pm\rangle = \frac{1}{\sqrt{2}} (|e\rangle \pm |g\rangle)\). The Hamiltonian $H_2$, in terms of the dressed-state basis $|\pm\rangle$, can be rewritten as
\begin{align}\label{4}	
	H_{3}=&\widetilde{\Omega}_{d}(\sigma_{++} -\sigma_{--})\notag\\
	&+\frac{1}{2}[g_a a e^{i(\delta-\delta_d)t}+g_b b e^{-i(\delta+\delta_d)t}](\sigma_{++}-\sigma_{--}-\sigma_{+-}+\sigma_{-+})\notag\\
	&+\frac{1}{2}[g_a a^{\dagger}e^{-i(\delta-\delta_d)t}+g_b b^{\dagger}e^{i(\delta+\delta_d)t}](\sigma_{++}-\sigma_{--}+\sigma_{+-}-\sigma_{-+}),
\end{align}
where $\sigma_{jk}=|j\rangle\langle k|$ ($j,k=+,-$), and $\widetilde{\Omega}_{d}=\sqrt{\Omega_{d}^{2}+(\frac{\delta_d}{2})^{2}}$. We choose $\delta_d\ll|\Omega_d|$ so that $\widetilde{\Omega}_{d}\simeq\Omega_d$. Then we work in the interaction picture by the transformation $U_{3}=e^{-i \Omega_{d}(\sigma_{++} -\sigma_{--}) t}$ , and the Hamiltonian becomes
\begin{align}\label{5}	
	H_{4}=&\frac{1}{2}\Big\{[g_{a}ae^{i\delta t}+g_{b}be^{-i\delta t}](\sigma_{++}-\sigma_{--}\notag\\
	&-\sigma_{+-}e^{i2\Omega_d t}+\sigma_{-+}e^{-2i\Omega_dt})e^{-i\delta_d t}+\rm H.c.\Big\}.
\end{align}
When $\frac{g_{a,b}}{2}\ll |\delta|,|\Omega_d|$, all terms in Eq.$~$(\ref{5}) can be regarded as high-frequency oscillation terms, and the effective Hamiltonian can be expressed as \cite{40a}
\begin{align}\label{6}	
	H_{\text{eff}} = -iH_4(t) \int H_4(t') \, dt'.
\end{align}
Substituting Eq.$~$(\ref{5}) into Eq.$~$(\ref{6}) and ignoring the fast oscillation terms. We find that, as expected, the following effective Hamiltonian can be obtained
\begin{align}\label{7}	
	&H_{\text{eff}} = H_0+H_{\text{int}},\notag\\
	&H_0=\frac{\Omega_d}{4\Omega_d^2-\delta^2} (g_a^2 a^\dagger a+g_b^2 b^\dagger b)(\sigma_{++}-\sigma_{--})\notag\\
	 &~~~~~~+\frac{2}{\delta}\sum\limits_{l=+,-}(\frac{g_a^2}{4}\frac{\delta+l\Omega_d}{\delta+2l\Omega_d}-\frac{g_b^2}{4}\frac{\delta-l\Omega_d}{\delta-2l\Omega_d})\sigma_{ll},\notag\\
	&H_{\text{int}}=\frac{\Omega_d g_a g_b}{\delta^2-4\Omega_d^2}ab(\sigma_{++}-\sigma_{--})e^{-2i\delta_d t}+\rm H.c..\notag\\
\end{align}
By performing the unitary transformation $U_4= e^{-i H_0 t}$ and neglecting the counter-rotating terms, we obtain the Hamiltonian in the interaction picture
\begin{align}\label{8}	
	H_{6} =& \frac{\Omega_{d}g_a g_b}{\delta^2-4\Omega_{d}^2}[(e^{-i[2\delta_d+\frac{\Omega_d}{4\Omega_d^2-\delta^2}(g_a^2+g_b^2)]t}\sigma_{++}ab\notag\\
	&-e^{i[-2\delta_d+\frac{\Omega_d}{4\Omega_d^2-\delta^2}(g_a^2+g_b^2)t]}\sigma_{--}ab)+\rm H.c.].
\end{align}
When the qubit is initially prepared in the state $|+\rangle$, and the driving frequency and strength satisfy the following relation
\begin{align}\label{90}	
	2\delta_d=\frac{\Omega_d}{\delta^2-4\Omega_d^2}(g_a^2+g_b^2),
\end{align}
we can obtain the parametric-amplification-type Hamiltonian for the two acoustic modes
\begin{align}\label{9}	
	H_7= \frac{\Omega_{d}g_a g_b}{\delta^2-4\Omega_{d}^2}(ab+a^{\dagger}b^{\dagger}).
\end{align}
For the case of the qubit being initially prepared in the state $|-\rangle$, the Hamiltonian above can also be obtained under the condition $2\delta_d=-\frac{\Omega_d}{\delta^2-4\Omega_d^2}(g_a^2+g_b^2)$. The Hamiltonian above indicates that the phonon modes $a$ and $b$ are coupled by the parametric down-converted interaction, hence the entanglement between them can be generated directly by the unitary evolution of the system. The physical explanation for the generation of entanglement is as follows: by engineering the system dynamics properly such that two acoustic modes create (or annihilate) a phonon simultaneously, so the phonon numbers in two modes cannot change independently. Their quantum states cannot be described separably using Fock-state basis, indicating the presence of quantum correlation \cite{41,42}.
	
\section{The quantitative results of the acoustic-mode entanglement}
We have qualitatively demonstrated the entanglement between two acoustic modes can be generated by engineering the parametric-amplification-type Hamiltonian in the section above. Now we quantitatively study the entanglement by solving the system dynamics equation using experimentally feasible parameters. When the dissipations and fluctuations are considered, the system dynamics can be described by the Lindblad master equation \cite{43,44}
\begin{align}\label{10}
	\frac{d}{dt}\rho=&-i[H,\rho]+\kappa_q(\overline{n}_{q}+1)\mathcal{L}_{\sigma_{ge}}\rho+\kappa_q\overline{n}_{q}\mathcal{L}_{\sigma_{eg}}\rho\notag\\
	&+\gamma_a(\overline{n}_a+1)\mathcal{L}_{a}\rho+\gamma_a\overline{n}_a\mathcal{L}_{a^{\dagger}}\rho\notag\\
	&+\gamma_b(\overline{n}_b+1)\mathcal{L}_{b}\rho+\gamma_b\overline{n}_b\mathcal{L}_{b^{\dagger}}\rho,
\end{align}
where $\mathcal{L}_{o}\rho=(o\rho o^{\dagger}-\frac{1}{2}o^{\dagger}o \rho-\frac{1}{2}\rho o^{\dagger}o)$ ($o=a,a^{\dagger},b,b^{\dagger},\sigma_{ge},\sigma_{eg}$). $\gamma_{a}$ ($\gamma_{b}$) describes the damping of the acoustic mode $a$ $(b)$ and $\kappa_q$ is the decay rate of the qubit. $\overline n_{j}={(e^{{\hbar\omega_j}/{k_{B}T}} - 1)}^{-1}$ ($j=a,b,q$) is the mean thermal occupation number with $T$ being the bath temperature. By solving the master equation, one can obtain the density operator of the cQAD system, and thus computing the expected values of the physical quantities. To explore the entanglement between two bosonic modes, we introduce the position and momentum quadratures of two acoustic modes
\begin{equation}
	\begin{aligned}\label{11}
		&X_a=\frac{ a+a^\dagger}{\sqrt{2}},\qquad X_b=\frac{ b+b^\dagger}{\sqrt{2}}, \\
		&Y_a=\frac{ a-a^\dagger}{i\sqrt{2}},\qquad Y_b=\frac{b-b^\dagger}{i\sqrt{2}}.\\
	\end{aligned}	
\end{equation}
For the system with linearity evolution initially in Gaussian states, its quantum properties can be entirely characterized by the covariance matrix (CM) $\sigma$ \cite{45}. By expressing the quadratures as the column vector $R=[X_{a}, Y_{a}, X_{b}, Y_{b}]^T$, the elements of the CM $\sigma_{m}$ for the two acoustic modes can be written as
\begin{align}\label{12}
	\sigma_{k,l}=\left\langle R_k R_l+R_l R_k\right\rangle/2,
\end{align}
where $k,l=1,2,3,4$. The two-mode $4\times4$ CM $\sigma_{m}$ can be written as the block matrix form
\begin{align}\label{13}
	\sigma_{m}=\left( \begin{array}{cccccc}
		V_a&&V_{ab}\\
		V_{ab}^T&&V_b
	\end{array} \right),
\end{align}
where $V_a$, $V_b$, and $V_{ab}$ are 2$\times$2 matrices. We adopt the logarithmic negativity $E_N$ \cite{46,47,48} to quantify the entanglement between two phonon modes, which is a reliable quantitative estimate of continuous-variable entanglement. The definition of $E_N$ is given by:
\begin{align}\label{14}
	E_N=\mathrm{max}[0,-\mathrm{ln}(2\eta^-)],
\end{align}
with $\eta^{-}\equiv2^{-1/2}$$\left\{\Sigma-[\Sigma^2-4 \mathrm{det}\sigma_{m}]^{1/2}\right\}^{1/2}$ and $\Sigma\equiv\mathrm{det}V_a+\mathrm{det}V_b-2\mathrm{det}V_{ab}$. If $E_N>0$, i.e., $\eta^-<1/2$, the phonon modes are entangled and the larger $E_N$ the higher the degree of the entanglement.
\begin{figure}
	\centering
	\includegraphics[height=6cm,width=8.5cm]{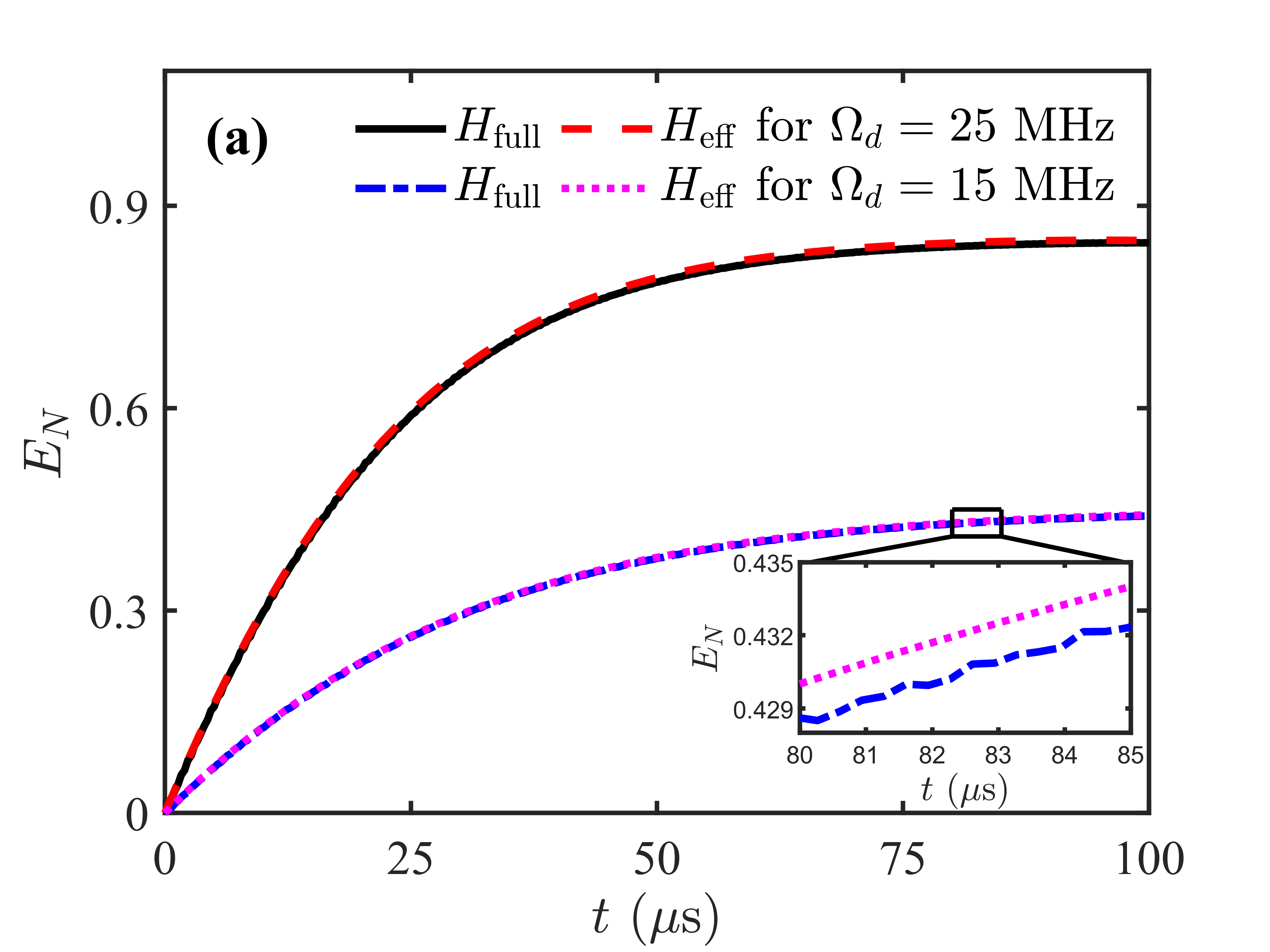}
	\includegraphics[height=6cm,width=8.5cm]{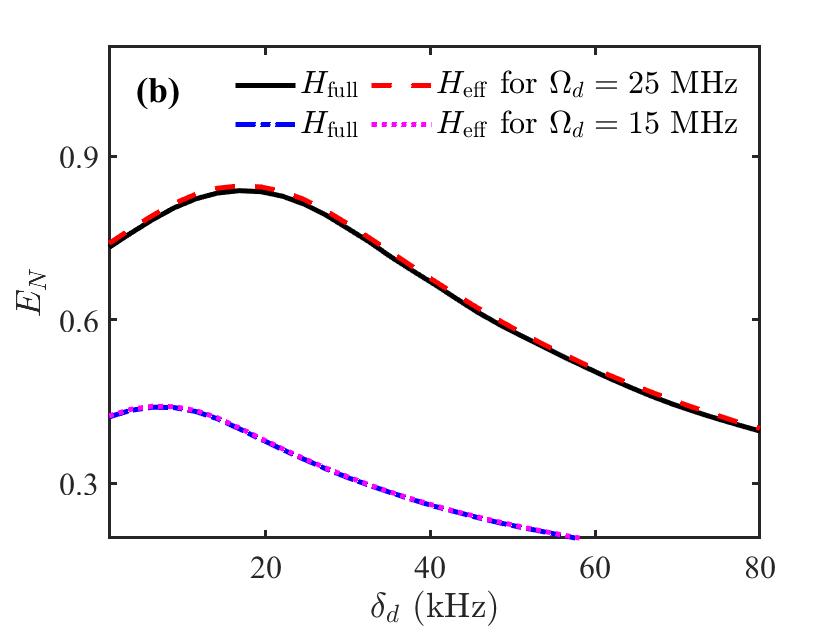}
	\caption{(a) The acoustic-mode entanglement $E_N$ as a function of time $t$ for different values of the driving amplitude $\Omega_d$. (b) The entanglement $E_N$ for the time $t=100$ $\mu s$ versus the detuning $\delta_d$ for different $\Omega_d$. See text for the other parameters.}
	\label{2}
\end{figure}
\begin{figure}
	\centering
	\includegraphics[height=6cm,width=8.5cm]{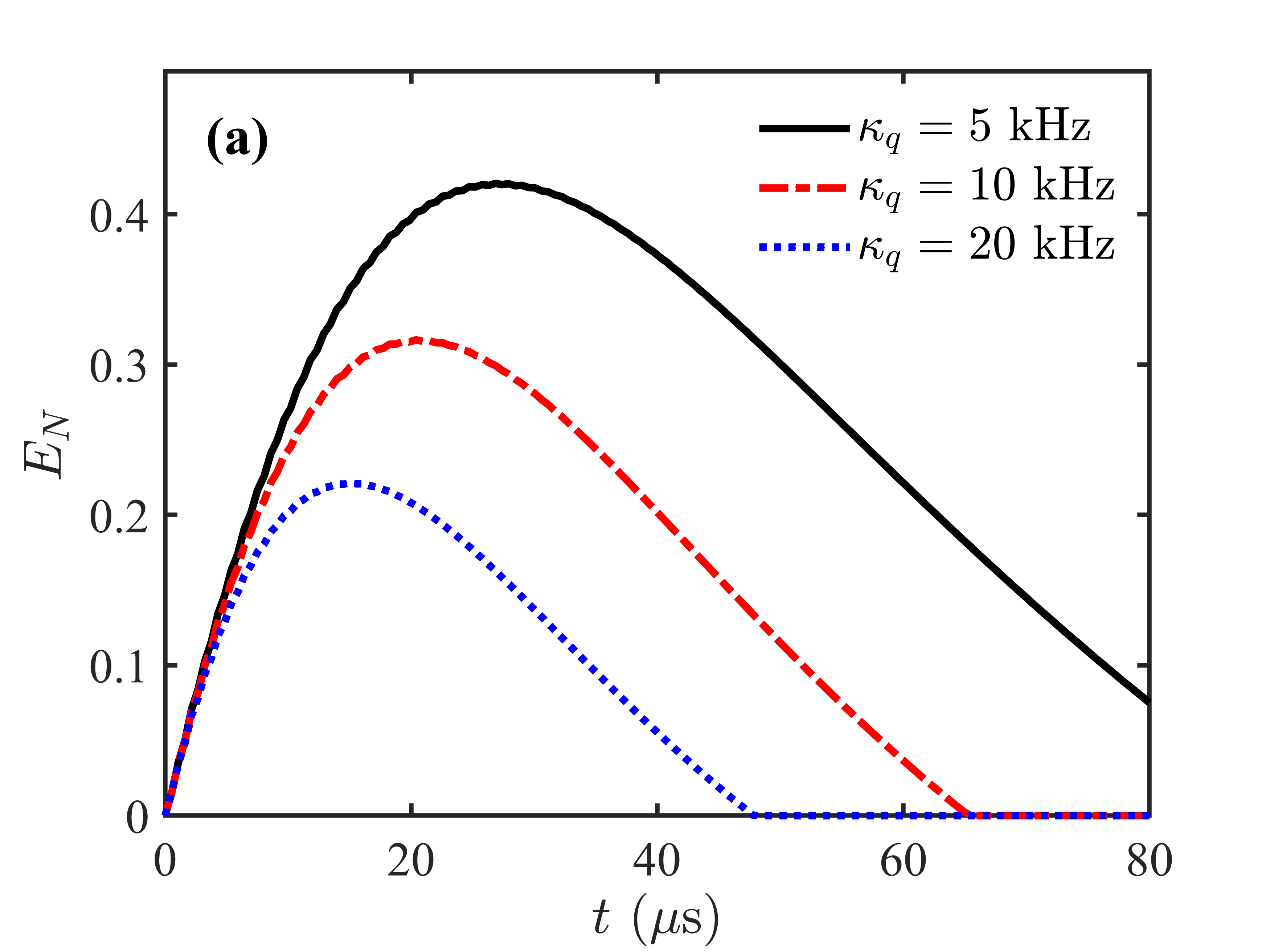}
	\includegraphics[height=6cm,width=8.5cm]{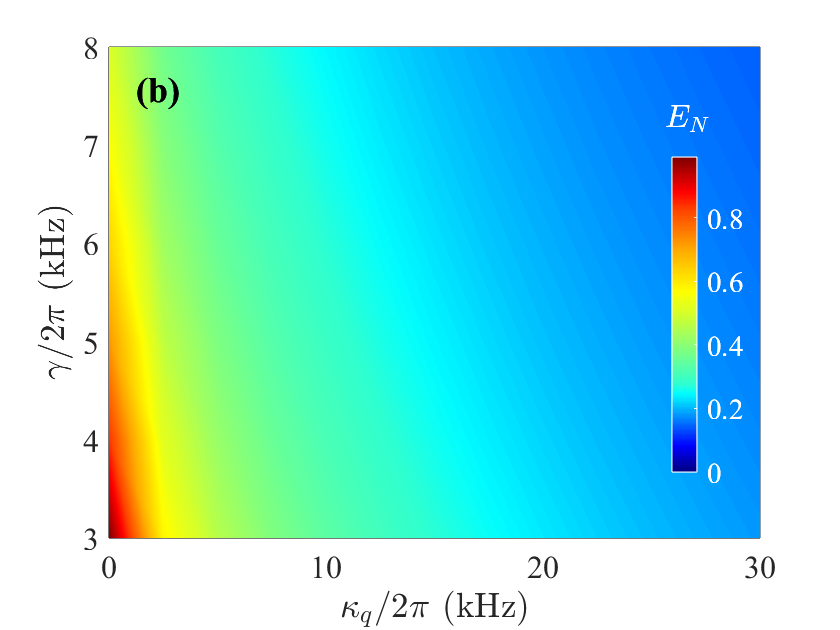}
	\caption{(a) The entanglement $E_N$ using the full Hamiltonian as functions of the time $t$ for different qubit decay rates $\kappa_q$, where $\Omega_d=25$ $\rm MHz$, $\gamma_a/2\pi=4.7 $ $\rm kHz $, and $\gamma_b/2\pi=3.3 $ $\rm kHz $. (b) The maximum entanglement $E_N$ versus $\kappa_q$ and $\gamma$ for $\Omega_d=25$ $\rm MHz$. Other parameters are the same as those in Fig.$~$\ref{2}.}
	\label{3}
\end{figure}	
\begin{figure}[t]
	\centering
	\includegraphics[height=6cm,width=8.5cm]{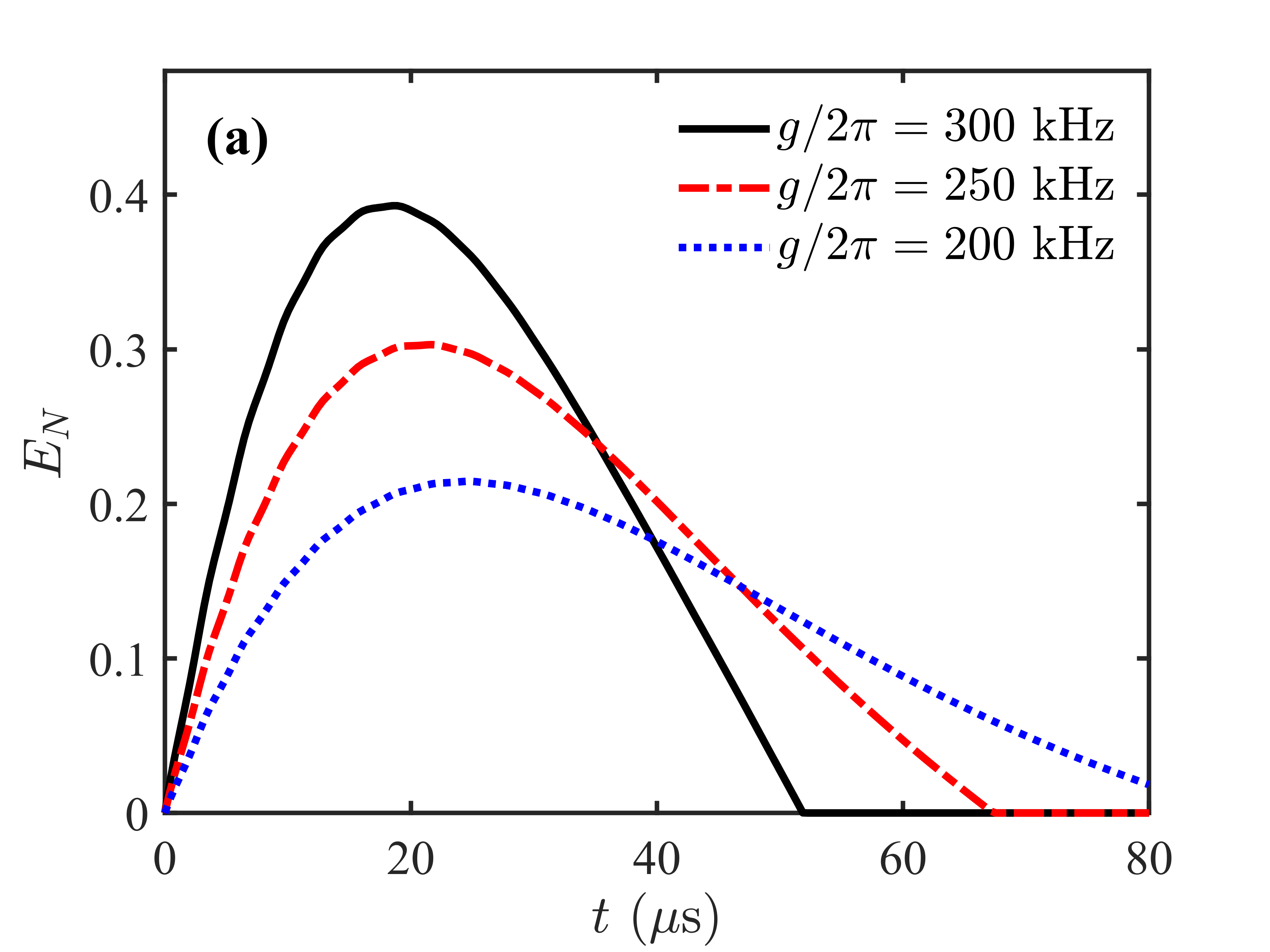}
	\includegraphics[height=6cm,width=8.5cm]{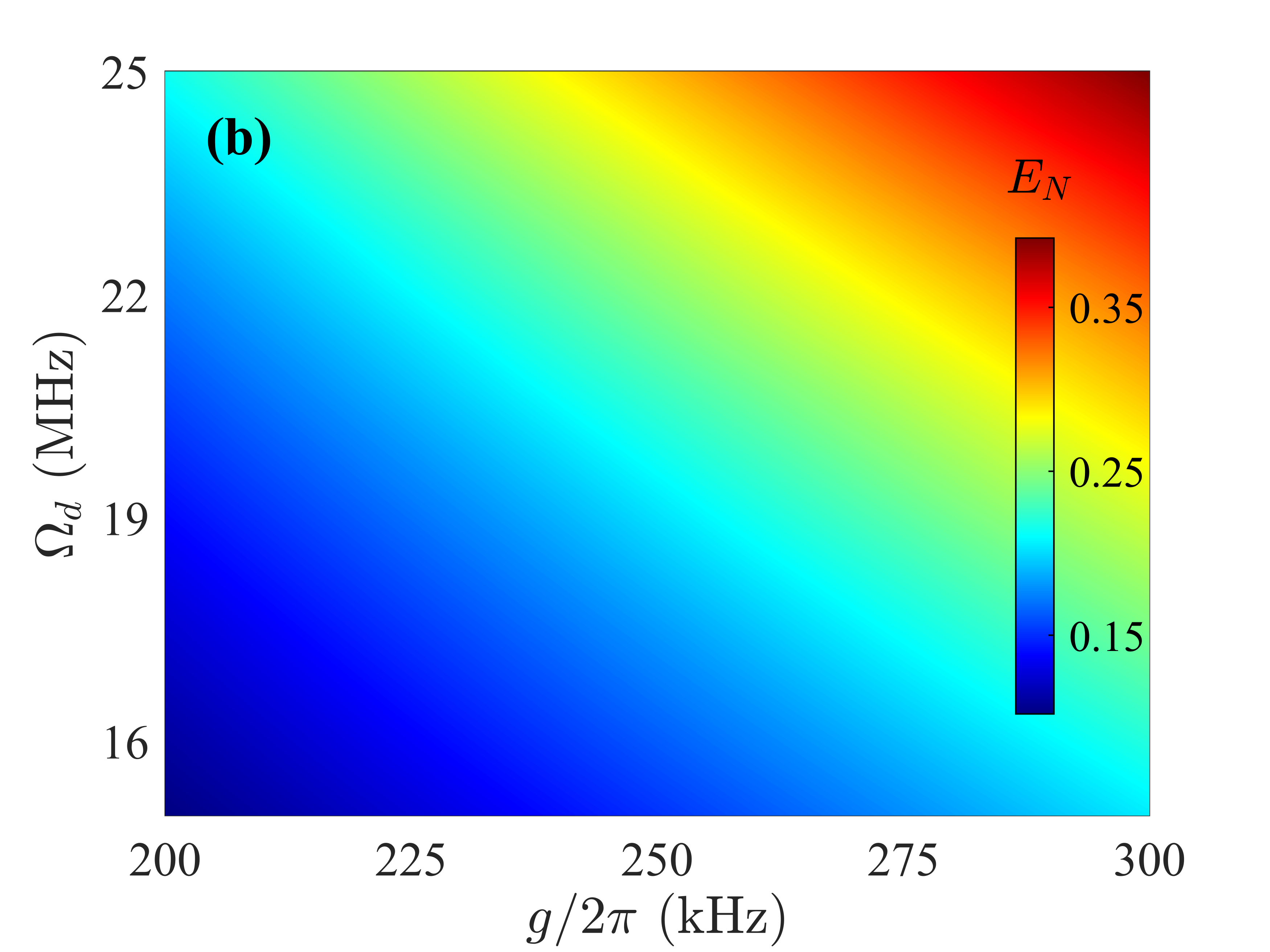}
	\caption{(a) The entanglement $E_N$ as a function of time $t$ with different values of the qubit-phonon coupling strength $g$ for $\Omega_d=25$ $\rm MHz$ and $\kappa_q/2\pi=10$ $\rm kHz$. (b) Density plot of the entanglement $E_N$ versus the coupling strength $ g $ and the driving amplitude $\Omega_d$ with $\kappa_q/2\pi=10$ $\rm kHz$. The other parameters are the same as those in Fig.$~$\ref{2}.}
	\label{4}
\end{figure}
\begin{figure}
	\includegraphics[height=6cm,width=8.5cm]{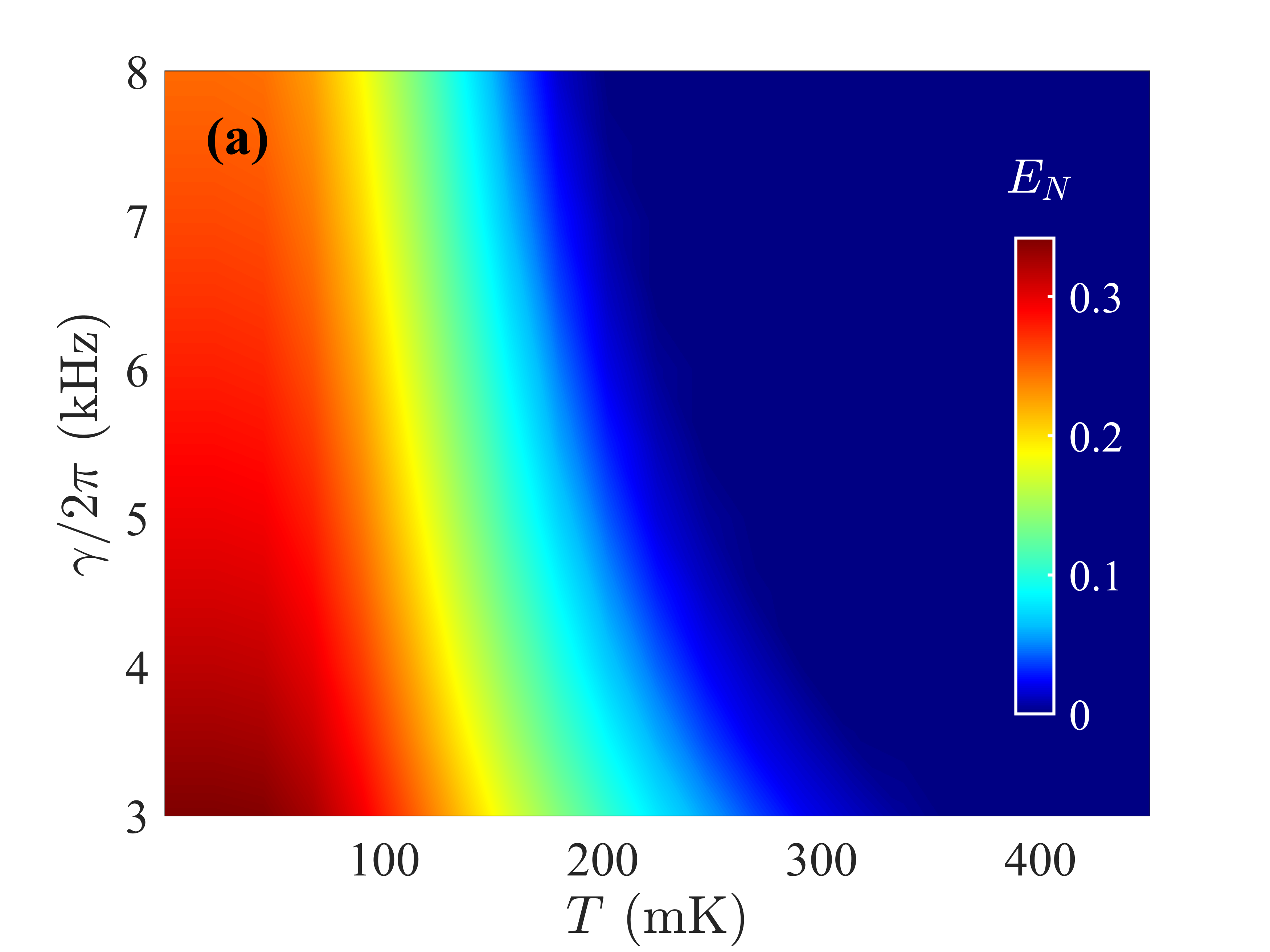}
	\includegraphics[height=6cm,width=8.5cm]{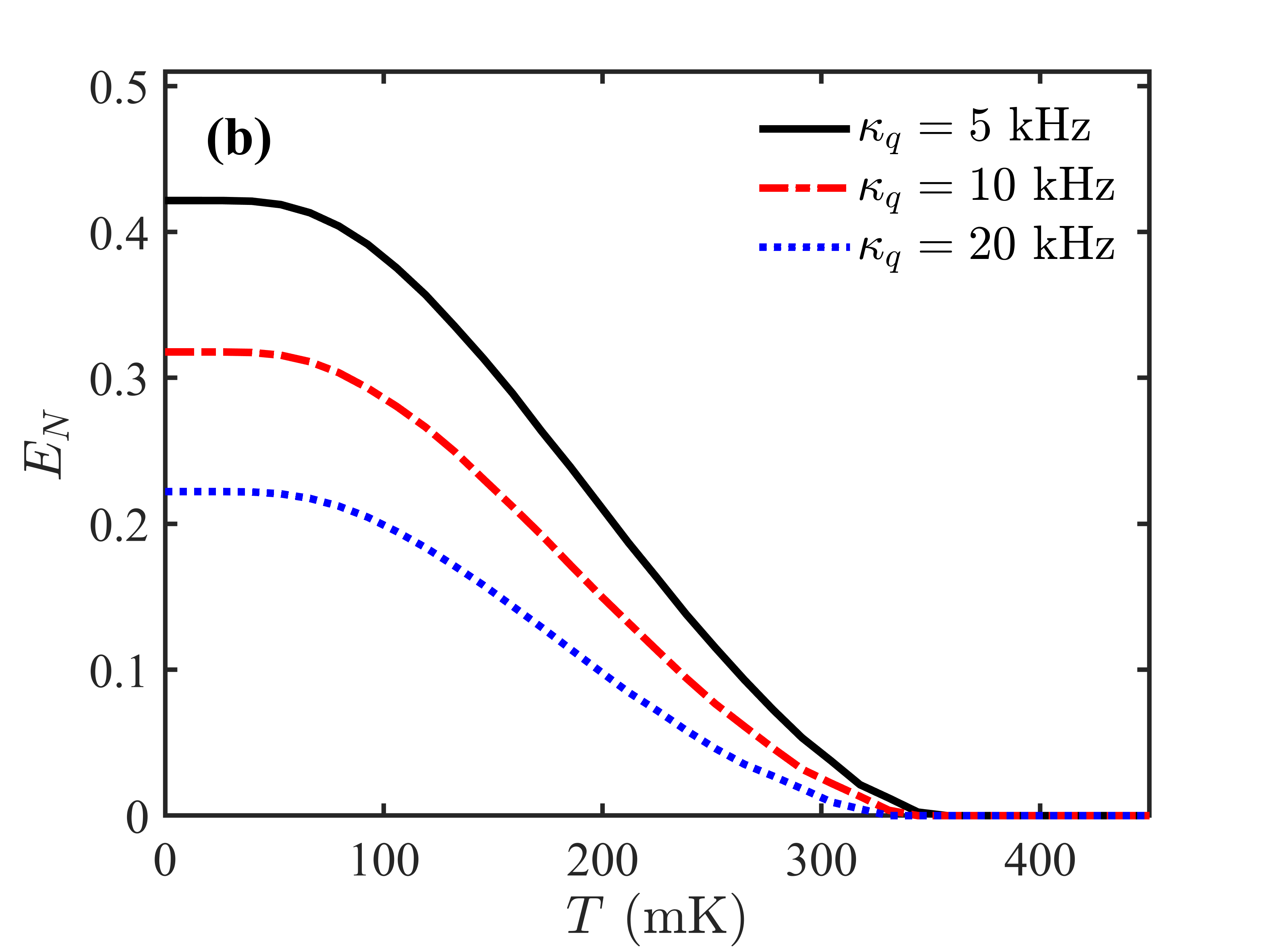}
	\caption{(a) Density plot of the maximum entanglement $E_N$ versus the phonon dissipation $\gamma$ and temperature $T$ for $\kappa_q/2\pi=10$ $\rm kHz$. (b) The maximum entanglement $E_N$ versus the temperature $T$ for different values of $\kappa_q$. We take $\Omega_d=25$ $\rm MHz$ and the other parameters are the same as those in Fig.$~$\ref{2}.}
	\label{5}
\end{figure}	
	
Now we discuss the time evolution of entanglement between two bosonic modes by numerically solving the master equation with the full Hamiltonian $(H_3)$ and the effective Hamiltonian $(H_7)$, respectively. The experimentally feasible parameters are selected \cite{29}: $\omega_a/2\pi=5.9236 $ $\rm GHz$, $\omega_b/2\pi=5.9488 $ $\rm GHz$, $g/2\pi=g_a/2\pi=g_b/2\pi=257 $ $\rm kHz $, $\gamma_a/2\pi=4.7 $ $\rm kHz $, and $\gamma_b/2\pi=3.3 $ $\rm kHz $. These parameters can also ensure the system is operated in the phonon-qubit dispersive coupling regime $g\ll |\omega_q-\omega_{a,b}|$. Here we temporarily assume that $\kappa_q=0$ and $T=50$ $\rm mK$. Initially, the phonon modes are in the vacuum states, and the qubit is prepared in the state $|+\rangle$.
	
The evolutions of entanglement over time $t$ for different driving amplitudes $\Omega_d$ are plotted in Fig.$~$\ref{2}(a), which shows that the considerable entanglement between two acoustic modes can be achieved after evolving for tens of microseconds. Moreover, the results obtained using the full Hamiltonian (solid and dash-dot lines) and the effective Hamiltonian (dashed and dotted lines) agree well with each other, indicating that the approximation used for deriving the effective Hamiltonian is very reasonable. Meanwhile, one can see the entanglement for $\Omega_d=25$ MHz is greater than that for $\Omega_d=15$ MHz, which can be explained through Eq.$~$(\ref{9}). In Eq.$~$(\ref{9}), the effective coupling strength is $G=\frac{\Omega_{d}g^2}{\delta^2-4\Omega_{d}^2}$. Obviously, $\Omega_d$ cannot equal to $|\delta|/2$ which causes the system to infinity, and the effective coupling strength will increase with increasing $\Omega_d$ for $\Omega_d<|\delta|/2$, leading to the increase of the entanglement. Nevertheless, the driving strength $\Omega_d$ and detuning $\delta_d$ cannot be arbitrarily chosen, because they are also constrained by Eq.$~$(\ref{90}). To discuss the effect of $\Omega_d$ and $\delta_d$ on the entanglement, we plot the entanglement as functions of the detuning $\delta_d$ for different $\Omega_d$ and $t=100$ $\mu s$ in Fig.$~$\ref{2}(b). One can see, for a given $\Omega_d$, there is an optimal $\delta_d$, which corresponds to Eq.$~$(\ref{90}) being satisfied. On the other hand, the entanglement can be present for a wide range of $\delta_d$, which means that the requirements for experimental parameters in this scheme are relatively relaxed, i.e., Eq.$~$(\ref{90}) does not need to be strictly satisfied.	
	
What should be emphasized is that, to demonstrate the validity of the effective Hamiltonian, we have chosen the decay rate of qubit $\kappa_q=0$ in Fig.$~$\ref{2}. In practice, $\kappa_q$ is bound to affect the performance of the scheme, so it must be taken into account. In Fig.$~$\ref{3}(a), we plot the entanglement dynamics for different $\kappa_q$ by using the full Hamiltonian (the results obtained using the effective Hamiltonian is not affected by $\kappa_q$). From Fig.$~$\ref{3}(a), we can see the obtained entanglement will decay to 0 with increasing the interaction time, and the smaller $\kappa_q$, the stronger  entanglement that can be obtained. That's because the acoustic modes will dissipate through the qubit due to the exchange interaction between them, although the dissipation rate of the acoustic mode is very small. Therefore, to make the entanglement continue for a long time, the interaction between acoustic modes and the qubit can be turned off when the entanglement reaches the maximum. In order to visually show the impact of system dissipations, we choose the two acoustic modes has the same dissipation rate $\gamma=\gamma_a=\gamma_b$, and plot the maximum entanglement versus $\kappa_q$ and $\gamma$ in Fig.$~$\ref{3}(b), which shows the degree of entanglement is significantly affected by $\kappa_q$. Fortunately, the qubit decay rate of $\kappa_q/2\pi\sim10$ kHz and the phonon dissipation of $3\sim4$ kHz have been experimentally achieved in cQAD systems~\cite{29,30}. Therefore, Fig.$~$\ref{3}(b) indicates the entanglement can be obtained within the range of experimentally feasible parameters, even if $\kappa_q$ reaches $2\pi\times30$ kHz.
	
Though the qubit is adiabatically eliminated in the derivation of the effective Hamiltonian, the qubit-phonon coupling is crucial for mediating the indirect coupling of two acoustic modes, which is reflected in Eq.$~$(\ref{9}) where the effective coupling strength is related to the driving amplitude $\Omega_d$ and the qubit-phonon coupling strength $g$. Therefore, we here investigate the dependence of $E_N$ on parameters $g$ and $\Omega_d$. We plot the evolution of entanglement over time under different qubit-phonon coupling strength $g$ for $\Omega_d=25$ $\rm MHz$ in Fig.$~$\ref{4}(a), which shows the maximum entanglement is significantly affected by $g$. Meanwhile, it can be found that the greater the coupling strength $g$, the faster entanglement between phonon modes decays to 0. That's because the stronger exchange interaction between qubit and phonon modes will result in the faster dissipating of phonon modes through the qubit. For clarity, we plot the maximum entanglement $E_N$ versus the coupling strength $g$ and the driving amplitude $\Omega_d$ in Fig.$~$\ref{4}(b). It can be seen that, similar to the effect of $\Omega_d$, the increase of $g$ also leads to the enhancement of $E_N$. This is attributed to the fact that the effective coupling strength $G$ is positively related to both $\Omega_d$ and $g$ in the appropriate range. Recent experiments have shown that the qubit-phonon coupling strength $g$ can reach about 300 kHz \cite{30}, so the presented scheme can generate considerable entanglement under current experimental conditions.
	
Finally, we discuss the influence of system dissipations and environmental temperature on the scheme comprehensively. Figure \ref{5}(a) shows the maximum entanglement versus phonon dissipation $\gamma$ and temperature $T$, which indicates the scheme has strong robustness. For example, the entanglement can be still obtained even when $T\sim$350 mK. We plot the maximum entanglement versus $T$ for different values of $\kappa_q$ in Fig.~\ref{5}(b) where $\gamma_a/2\pi=4.7$ $\rm kHz$, and $\gamma_b/2\pi=3.3$ $\rm kHz $. We can see that, though the qubit decay rate will reduce the degree of entanglement significantly, it has little influence on the robustness of entanglement against temperature.
	
\section{Conclusions}
	
We have proposed a scheme for generating the entanglement between two long-lived acoustic modes by engineering their parametric-amplification-type Hamiltonian in cQAD system, where the qubit is only driven by a single microwave field and couples to the two acoustic modes dispersively ($g\ll |\omega_q-\omega_{a,b}|$). The indirect interaction between the two acoustic modes can be mediated by adiabatically eliminating the qubit. We demonstrated that, by appropriately choosing the drive frequency and strength, the parametric-amplification-type Hamiltonian between two acoustic modes can be established. Therefore, the entanglement can be straightforward generated by the unitary evolution of the system. Through comparing the results obtained from the full Hamiltonian and the effective Hamiltonian, it has been confirmed that the derived effective Hamiltonian is a very good approximation. We numerically analyzed the dependence of the scheme on the system parameters in detail, and the results showed that the scheme can be achieved under the recent available experimental conditions \cite{29,30}. Furthermore, the robustness of the scheme against the system dissipations and environmental temperature has been demonstrated. Therefore, this work provides a simple and feasible approach to generate entanglement in cQAD system, and may be meaningful for the quantum information processing based on the solid-state mechanical resonator.
	
\begin{center}$\mathbf{Acknowledgments}$\end{center}
	
This work is supported by the Innovation Program for Quantum Science and Technology Grant No. 2023ZD0300400, the National Key Research and Development Program of China under Grant No. 2021YFA1402002, and the National Natural Science Foundation of China under Grant Nos. 12274274, U21A6006, and U21A20433.

\end{document}